\pdfoutput=1

\documentclass[prl,twocolumn,showpacs,preprintnumbers,amsmath,amssymb,
superscriptaddress,nofootinbib]{revtex4-1}
\usepackage{epsfig}
\usepackage{graphicx}
\usepackage{cancel}
\usepackage{color}

\newcommand{\Ddots}{\hbox to 1em{.\hss.\hss.\hss}}

\begin{document}

\preprint{}

\title{Recursion Relations for One-Loop Goldstone Boson Amplitudes}

\author{Christoph Bartsch}
\affiliation{Institute of Particle and Nuclear Physics, Charles University, Prague, Czech Republic}

\author{Karol Kampf}
\affiliation{Institute of Particle and Nuclear Physics, Charles University, Prague, Czech Republic}

\author{Jaroslav Trnka}

\affiliation{Institute of Particle and Nuclear Physics, Charles University, Prague, Czech Republic}

\affiliation{Center for Quantum Mathematics and Physics (QMAP), University of California, Davis, CA, USA}

%\date{\today}

\begin{abstract}
In this letter, we construct the recursion relations for one-loop planar integrands in the $SU(N)$ non-linear sigma model. This generalizes the soft recursions for tree-level amplitudes in a variety of quantum field theories with special soft limits. The main ingredient is the definition of the one-loop planar integrand, which is fixed by cuts in the sense of generalized unitarity and by requiring the Adler zero on all external legs. We show that this does not uniquely fix the integrand, so additional constraints on the soft behavior of the loop momentum have to be imposed. Our work is the first step in extending modern amplitudes methods for loop amplitudes to effective field theories with special soft limits.

\end{abstract}

\maketitle

%%%%%%%%%%%%%%%%%
\section{Introduction}
%%%%%%%%%%%%%%%%%

One of the major advances of the modern S-matrix program is the construction of scattering amplitudes from their physical properties without the use of the standard Feynman diagram prescription. At tree-level, scattering amplitudes in a large class of quantum field theories are constructible from factorizations via on-shell recursion relations \cite{Britto:2004ap,Britto:2005fq}. Discovered in the context of gauge theory, on-shell recursion relations have been extended to a large class of theories \cite{Cheung:2015cba}. In all cases, the tree-level S-matrix was completely fixed by factorizations. 

In \cite{Cheung:2015ota,Cheung:2016drk}, two of the authors formulated the \emph{soft recursion relations} which use a special soft limit behavior of the amplitude as an additional input to completely fix the tree-level amplitude. These special soft theories include the $SU(N)$ non-linear sigma model (NLSM), Dirac-Born-Infeld theory, or Galileons. In all cases, the scattering amplitudes vanish as $A_n={\cal O}(p^\sigma)$ for $p\rightarrow0$, where the integer $\sigma$ denotes the degree of the soft limit. In \cite{Cheung:2018oki, Elvang:2018dco}, these methods were further generalized to vector field theories and supersymmetric theories, and in \cite{Kampf:2019mcd, Luo:2015tat} to theories with non-vanishing (but known) soft limits. In all cases, the tree-level amplitudes were completely fixed by factorizations and the behavior in the soft limit. 
%The recursion relations provided a practical tool to reconstruct them from this input data.

In a parallel line of research, the loop recursion relations were constructed for amplitudes in the planar maximally supersymmetric Yang-Mills (${\cal N}=4$ SYM) theory \cite{Arkani-Hamed:2010zjl}. The crucial ingredient is a unique definition of the \emph{planar loop integrand}, which is a rational function of external and loop momenta completely fixed by its singularities. These singularities correspond either to tree-level factorizations or loop cuts. It was shown in \cite{Caron-Huot:2010fvq} that the single cut of the $n$-point ${\cal N}=4$ SYM $\ell$-loop integrand is equal to the forward limit of the $n{+}2$-point $\ell{-}1$-loop integrand. This information is a sufficient input into the recursion relations which can be used to construct the planar ${\cal N}=4$ SYM integrand at arbitrary multiplicity and loop order from elementary tree-level amplitudes. 

While the tree-level recursion relations work for a large class of theories, the loop recursion is so far specific to this one particular theory. The reason is two-fold: only in planar (large $N$) theories we can define global loop variables and talk about the loop integrand as a single object (rather than a sum of diagrams); and in most theories the single cut of the loop integrand is divergent and needs to be regulated \cite{Benincasa:2016awv}.

In this letter, we study the one-loop amplitudes of Goldstone boson scattering processes in the NLSM. Based on conventional methods, only the four-point amplitude is known beyond the tree-level in the $SU(N)$ model, and the six-point one-loop result was calculated only recently \cite{Bijnens:2021hpq} for the $O(N)$ non-linear sigma model.

We focus on the planar integrand in the large $N$ limit of $SU(N)$ and construct it using unitarity methods. While most terms are fixed by standard cuts, we show that there is an ambiguity in tadpole terms. These terms integrate to zero in the dimensional regularization but are nevertheless important for the unique definition of the integrand. We fix this ambiguity using special soft limit constraints. Having defined the unique one-loop integrand, we construct the loop generalization of the soft recursion relations, which can be used to calculate the integrand to any number of points.

%%%%%%%%%%%%%%%%%
\section{Tree-level soft recursion relations}
%%%%%%%%%%%%%%%%%

We consider scattering amplitudes of Goldstone bosons in the $SU(N)$ non-linear sigma model. The theory can be parameterized by a Lagrangian 
\begin{align}
    \mathcal{L}_2 = \frac{F^2}{4} \langle \partial_\mu U \partial^\mu U^{-1} \rangle,\; U(x) = \sum_{k=0}^\infty u_k\Bigl(\frac{i\sqrt{2}}{F}\phi(x) \Bigr)^{\!k}\,,
    \label{Lagr}
\end{align}
where $\phi(x)=\phi^a(x) t^a$, with $t^a$ the generators of $SU(N)$ and $\langle \Ddots \rangle \equiv {\rm Tr}(\Ddots)$. We will use a general parametrization for $U(x)$, but one can always pick a familiar exponential form $u_k=1/k!$. At tree-level, we can write the $n$-point amplitude ${\cal A}_n$ as a sum over \emph{flavor-ordered amplitudes},
\begin{equation}
    {\cal A}_n = \!\!\!\sum_{\sigma\in S_n/Z_n}\!\!\! \frac{\langle t^{a_{\sigma(1)}}\Ddots t^{a_{\sigma(n)}}\!\rangle}{(2F^2)^{n/2-1}}\,A_n(p_{\sigma (1)},\Ddots,p_{\sigma (n)}), \label{flavor}
\end{equation}
analogous to color ordering in the Yang-Mills amplitudes. The ordered amplitude $A_n$ vanishes in the soft limit \cite{Kampf:2013vha}, 
\begin{equation}
   \lim_{p_k\rightarrow 0}A_n(p_1,\Ddots,p_n) = 0,   \label{soft}
\end{equation}
and makes manifest the Adler zero of the complete tree-level amplitude ${\cal A}_n$. Furthermore, the amplitude $A_n$ is completely fixed by factorization condition on all poles,
\begin{equation}
    A_n \xrightarrow[]{P^2=0} -A_L\frac{1}{P^2}A_R, \label{unit}
\end{equation}
for all $P=p_i{+}{\dots}{+}p_j$ and the soft limit condition (\ref{soft}). This allows us to write the \emph{soft recursion relations}. We first shift all momenta,
\begin{equation}
    \widehat{p}_{n} = p_{n} + z q_{n}\,,\quad
    \widehat{p}_{1} = p_{1} + z q_{1}\,,\quad\widehat{p}_k = p_k(1-a_kz)\,, \label{shift}
\end{equation}
where $k=2,{\dots},n{-}1$ and two momenta $q_n,q_1$ are fixed by on-shell conditions on shifted $\widehat{p}_{n}^2 = \widehat{p}_{1}^2 = 0$ and momentum conservation. The shifted amplitude $A_n(z)$ factorizes on poles $P_I^2(z_I^{\pm})=0$ into a product of sub-amplitudes (\ref{unit}), and vanishes in the soft limit $z=1/a_k$. We can use the residue theorem, 
\begin{equation}
    \oint \frac{dz}{z}\frac{A_n(z)}{F(z)} = 0 \label{GRT},
\end{equation}
where $F(z)=\prod_k (1-za_k)$, to reconstruct the original amplitude (residue on $z=0$) from the factorization poles of $A_n(z)$. Note that there are no poles at $1-za_k=0$ due to the zeroes of (\ref{soft}) and the pole at $z\rightarrow\infty$ in $A_n(z)$ is canceled by the insertion of the denominator factor of (\ref{GRT}). The tree-level amplitude $A_n$ can then be written as 
\begin{equation}
    A_n = \sum_{I,\pm} \mathop{\mathrm{Res}}_{z=z_I^{\pm}} \frac{A_L(z)A_R(z)}{z P_I^2(z)F(z)},
\end{equation}
where we sum over all factorization channels $P_I^2(z_I^{\pm})=0$.

%%%%%%%%%%%%%%%%%
\section{One-loop Feynman integrand}
%%%%%%%%%%%%%%%%%

Our goal is to define a unique NLSM \emph{loop integrand} and reconstruct it using the recursion relations. First, we realize that beyond tree-level the simple flavor ordered formula (\ref{flavor}) does not work because of the presence of multiple trace terms. However, in the large $N$ (planar) limit, the single trace dominates and we can write,
\begin{equation}
    {\cal A}_n^{\rm 1-loop} = \!\!\!\!\!\sum_{\sigma\in S_n/Z_n}\!\!\!\!\! \frac{\langle t^{a_{\sigma(1)}}\Ddots t^{a_{\sigma(n)}}\!\rangle}{(2F^2)^{n/2}}A^{\rm 1-loop}_n(p_{\sigma (1)},\Ddots,p_{\sigma (n)}).
\end{equation}
The ordered amplitude $A^{\rm 1-loop}_n(p_1,\dots,p_n)$ is UV divergent and needs to be regulated. The loop integrand ${\cal I}_n^{\rm 1-loop}$ is a rational function of the planar loop momentum $\ell$ and external momenta, 
\begin{equation}
    A^{\rm 1-loop}_n = \! \int d^4\ell\,\,{\cal I}_n^{\rm 1-loop}(\ell,p_k)\,.
\end{equation}
The loop momentum $\ell$ is uniquely defined (see \cite{Arkani-Hamed:2010pyv} for the discussion of dual variables) as the loop momentum flow between external legs $n$ and $1$. This gives a unique way of labeling all contributing Feynman diagrams. The loop integrand ${\cal I}_n$ is then given as a sum of appropriately labeled Feynman diagrams prior to the loop integration.

At four points, the contributing Feynman diagrams are 
\begin{align*}
\begin{split}
    {\cal{I}}_4^{\rm FD} &= \begin{matrix}\includegraphics[width=2.3cm]{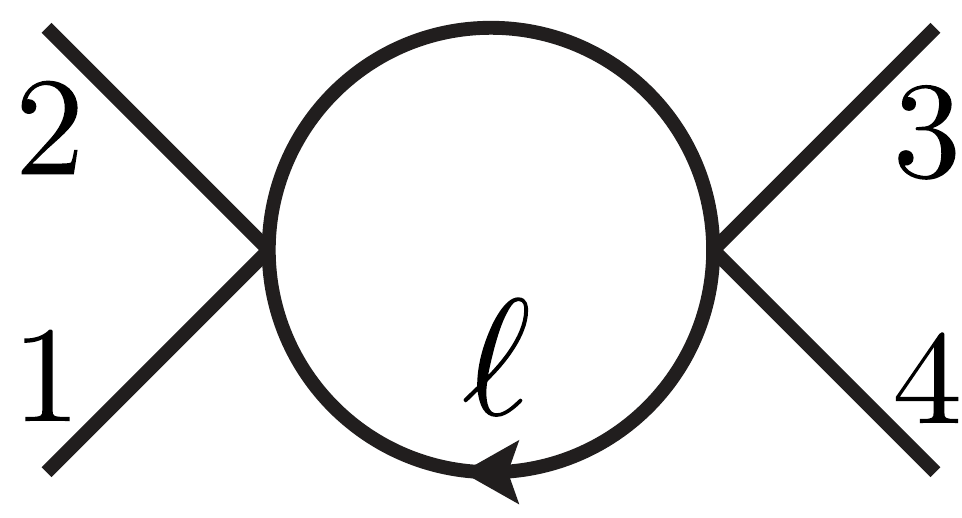}\end{matrix} + \begin{matrix}\includegraphics[width=2.3cm]{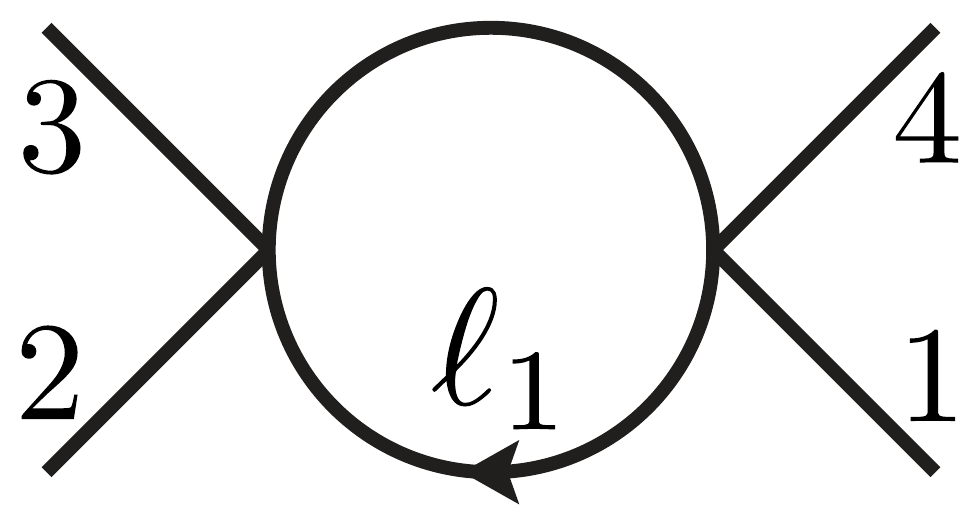}\end{matrix}\\ &+ \begin{matrix}\includegraphics[width=1.5cm]{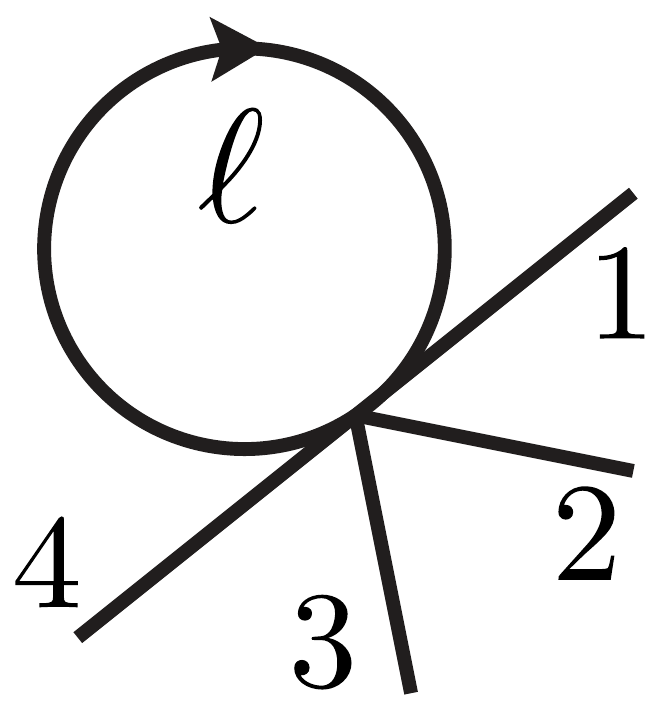}\end{matrix} + \begin{matrix}\includegraphics[width=1.5cm]{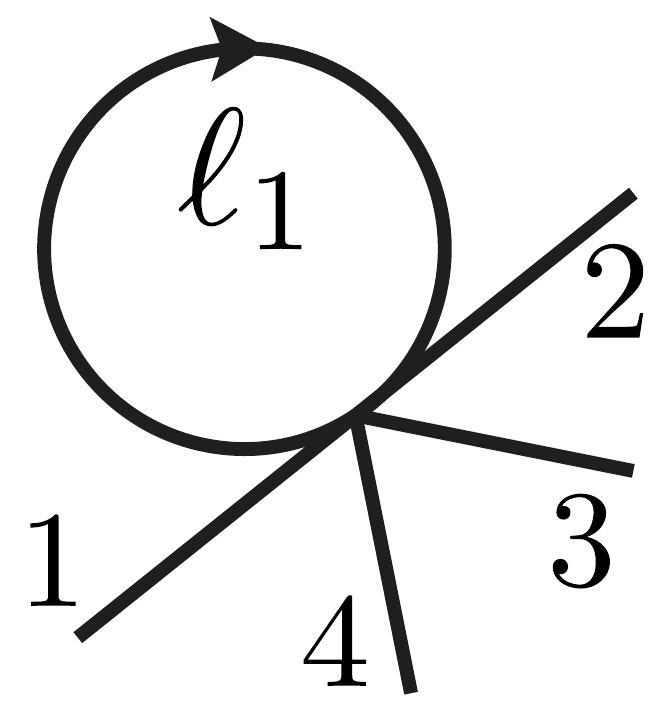}\end{matrix} + \begin{matrix}\includegraphics[width=1.5cm]{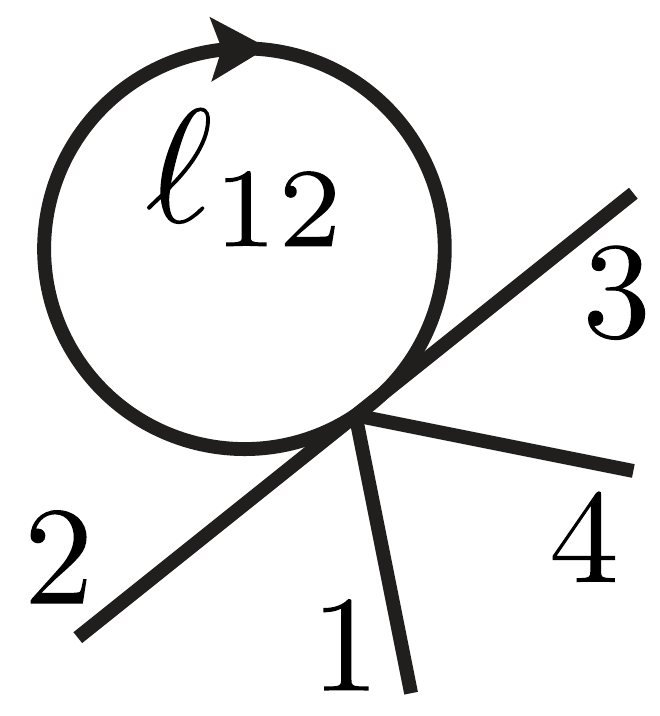}\end{matrix} +\begin{matrix}\includegraphics[width=1.5cm]{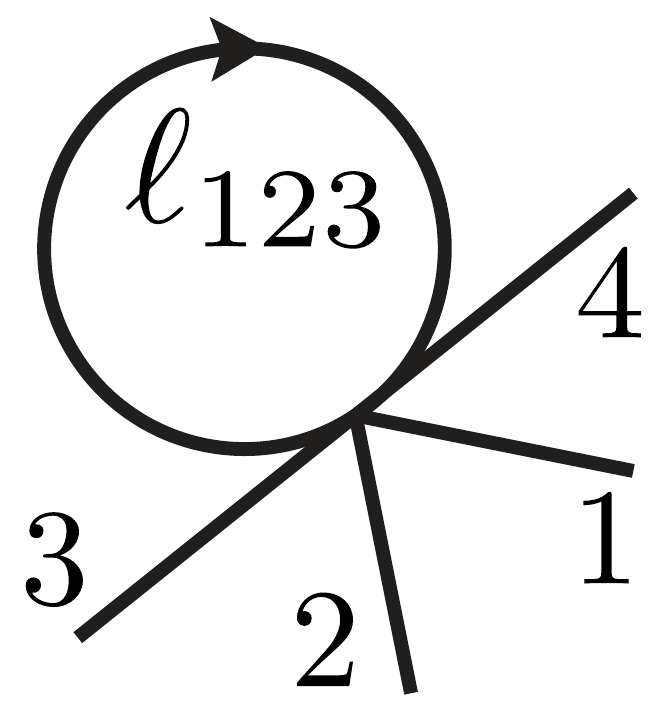}\end{matrix}\, ,
\end{split}
\end{align*}
where we defined $\ell_1\!=\!\ell +p_1$, $\ell_{12} \!=\! \ell + p_1 + p_2$, $\ell_{123} \!=\! \ell  + p_1  + p_2 + p_3$, and identify $\ell_{1234} \!\equiv\! \ell $ due to momentum conservation.
Using the stripped Feynman rules for ordered vertices $V_n(p_1,\Ddots,p_n)$ \cite{Kampf:2013vha},
\begin{align}
\begin{split}
    V_4 &= \frac{1}{2}s_{12}-4u_3 p_1^2 + \text{cyc},\\
    V_6 &= (1-8u_3)s_{12} + 8u_3^2 s_{123} + 16u_5p_1^2 + \text{cyc},
\end{split}
\end{align}
we evaluate the diagrams and get the Feynman integrand,
\begin{align}
     {\cal I}_4^{\rm FD}(l,p_j) &= 16u_3^2 - 32u_5 + \frac{n_4^{\rm FD}}{\ell^2} + \frac{n_{42}}{\ell^2 \ell_{12}^2} +\text{cyc}\,. \label{FDint}
\end{align}
Here we define for any function $f(\ell,p_1,\Ddots,p_n)$,
\begin{align*}
 f(\ell,p_1,\Ddots,p_n) + \text{cyc} \equiv {\textstyle\sum_{i=1}^n} f(\ell_{1\dots i},p_{1+i},\Ddots,p_{n+i}),
\end{align*}
with all indices understood modulo $n$. The numerators in (\ref{FDint}) are
\begin{align*}
    n_4^{\rm FD} &= (4u_3-1)(2s_{12}+\ell_1^2+\ell_{123}^2) + (8u_3-1)s_{23},\\
    n_{42} &= \frac12(s_{12} + \ell_1^2)(s_{12} + \ell_{123}^2).
\end{align*}
Thus the loop integrand obtained from Feynman diagrams does depend on a particular parametrization of the Lagrangian (i.e. parameters $u_3$, $u_5$). However, the amplitude $A_n^{\rm 1-loop}$ is independent of $u_3$, $u_5$, so terms involving these coefficients must integrate to zero and we can fix $u_3$, $u_5$ in ${\cal I}_4^{\rm FD}$ to any values, not changing the physical result for $A_n^{\rm 1-loop}$.

Let us try to fix the coefficients $u_3$, $u_5$ in ${\cal I}_4^{\rm FD}$ by imposing an additional constraint. The natural candidate is imposing the vanishing soft limit in any of the external momenta $p_k$,
\begin{equation}
    \lim_{p_k\rightarrow 0} {\cal I}_4^{\rm 1-loop}(\ell,p_j) = 0. \label{loopsoft}
\end{equation}
Taking the soft limit of (\ref{FDint}), e.g. on line $p_4$,
\begin{align}
    &\lim_{p_4\rightarrow 0} {\cal I}_4^{\rm FD}(\ell,p_j) = 2(32u_3^2 - 64u_5 + 4u_3 -1)\\ &+ (4u_3-1)\Bigl( \frac{\ell_1^2}{\ell^2} + \frac{\ell^2}{\ell_1^2} + \frac{\ell_{12}^2}{\ell^2} + \frac{\ell^2}{\ell_{12}^2} \Bigr) +4u_3\Bigl(\frac{\ell_1^2}{\ell_{12}^2} + \frac{\ell_{12}^2}{\ell_1^2}\Bigr),\nonumber
\end{align}
we see that for no values of $u_3$, $u_5$ the soft limit can be set to zero. We can conclude that the Feynman integrand ${\cal I}_4^{FD}$ does not vanish in the soft limit.

%%%%%%%%%%%%%%%%%
\section{Soft integrand}
%%%%%%%%%%%%%%%%%

The loop integrand is not a physical quantity, so we can make an arbitrary change in the Feynman integrand ${\cal I}_4^{\rm FD}$ as long as it integrates to the same function. In other words, we are free to add terms which integrate to zero. The only terms in (\ref{FDint}) which are fixed and can not be changed are the bubble diagrams, as they must reproduce a physical unitarity cut which is encoded in the structure of logarithms after integration. The tadpole (and constant) terms in (\ref{FDint}) can be changed in an arbitrary way as the corresponding integrals integrate to zero. Therefore, we start with the following ansatz,
\begin{equation}
    {\cal I}_4^{\rm ans} = \frac{\alpha_0}{4} + \frac{n_4^{\rm ans}}{\ell^2} + \frac{n_{42}}{\ell^2\ell_{12}^2} + \text{cyc},
\end{equation}
where the kinematical ansatz for $n_4^{\rm ans}$ has 5 independent constants, 
\begin{equation}
    n_4^{\rm ans} = \alpha_1 s_{12} + \alpha_2 s_{23} +\alpha_3 \ell_1^2 + \alpha_4 \ell_{12}^2 + \alpha_5 \ell_{123}^2 \label{ansatz}.
\end{equation}
Now imposing the soft limit constraint (\ref{loopsoft}) we fix
\begin{equation}
    \alpha_0 = 2,\,\,\alpha_3 = -1,\,\,\alpha_4 = 1,\,\,\alpha_5 = -1,
\end{equation}
while coefficients $\alpha_1$, $\alpha_2$ remain unfixed. This makes perfect sense because for $p_k\rightarrow 0$ (for any $k$) both Mandelstam invariants $s_{12}$ and $s_{23}$ are automatically zero, and therefore we can not get any constraints on $\alpha_1$, $\alpha_2$ coefficients. As a result, we get a two-parametric soft integrand ${\cal I}_4^{\rm S}(\alpha_1,\alpha_2)$. While this integrand is not (yet) uniquely fixed, we cannot obtain it from Feynman diagrams in any parametrization (\ref{Lagr}) of the Lagrangian. 

Next, we evaluate our integrand ${\cal I}_4^{\rm S}(\alpha_1,\alpha_2)$ on a single cut $\ell^2=0$,
\begin{align}
{\rm Cut}[{\cal I}_4^{\rm S}] &\equiv - B_6(p_1,p_2,p_3,p_4,-\ell,\ell) \label{cut2}\\ 
&= \alpha_1 s_{12} + \alpha_2 s_{23} - \ell_1^2 + \ell_{12}^2 - \ell_{123}^2  + \frac{2n_{42}}{\ell_{12}^2}. \nonumber
\end{align}
Naively, we can try to identify the on-shell function on the right-hand side with the forward limit of the six-point tree-level amplitude (as in the ${\cal N}=4$ SYM theory) but a careful inspection reveals that it is not the case. Nevertheless, it is an interesting function which can be constrained and used to fix $\alpha_1$, $\alpha_2$. In particular, we can impose that $B_6$ vanishes in the soft limit $\ell\rightarrow0$,
\begin{equation}
    \lim_{\ell\rightarrow0} B_6 = -(\alpha_1+2)s_{12} - \alpha_2 s_{23} \overset{!}{=} 0, \label{b6Sft}
\end{equation}
which fixes $\alpha_1=-2$, $\alpha_2=0$. Interestingly, $B_6$ also vanishes in the soft limits $p_2\rightarrow0$ and $p_3\rightarrow0$ but not in the limits $p_1\rightarrow0$ and $p_4\rightarrow0$ (if it did, it would be indeed the forward limit of the tree amplitude). Hence, we will call $B_6$ a \emph{half-soft on-shell function}. 

As a result, we get a unique 4-point soft integrand 
\begin{equation}
    {\cal I}_4^{\rm S} = \frac{1}{2} + \frac{n_4^{\rm S}}{\ell^2} +\frac{n_{42}}{\ell^2\ell_{12}^2} + \text{cyc},
    \label{I4s}
\end{equation}
where $n_4^{\rm S} = -2s_{12} - \ell_1^2 + \ell_{12}^2 - \ell_{123}^2$.

The generalization to $n$ points is straightforward. We expand the integrand
\begin{equation}
    {\cal I}_n^{\rm 1-loop} = \beta_{n,0} + {\textstyle\sum_k} \beta_{n,k}\,{\cal B}_n^{(k)} \label{npoint},
\end{equation}
where ${\cal B}_n$ denotes the standard basis of scalar loop integrands\cite{Bourjaily:2020qca},
\begin{align}
   {\cal{B}}_n = \begin{Bmatrix}\includegraphics[width=5.8cm]{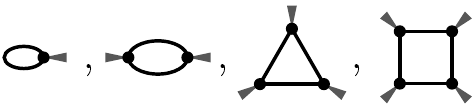}\end{Bmatrix},
\end{align}
with each of the above topologies representing multiple terms of the integrand with different numerators. All coefficients for boxes, triangles and bubbles are fixed by cuts in the framework of generalized unitarity \cite{Bern:1994cg}. The particular coefficients $\beta_{n,k}$ depend on the choice of basis ${\cal B}_n$ (choice of  numerators). Only tadpole terms ($k=1$) and the coefficient $\beta_{n,0}$ remain unfixed by this procedure. 

The coefficients $\beta_{n,k}$ are tree-like objects (sums of products of trees) with poles in external kinematics. The soft integrand must factorize on these poles into a lower point one-loop integrand and a tree-level amplitude
\begin{equation}
    {\cal I}_n^{\rm 1-loop} \xrightarrow[]{P^2=0}  -{\cal I}_L^{\rm 1-loop} \frac{1}{P^2} A_R^{\rm tree}.  \label{facttree}
\end{equation}
For example, the 6-point integrand ${\cal I}_6^{\rm 1-loop}$ factorizes on the pole $s_{123}=0$ as
\begin{align}
&-{\cal I}_4^{\rm 1-loop}(\ell,p_1,p_2,p_3,-P)\frac{1}{s_{123}} A_4^{\rm tree}(P,p_4,p_5,p_6)\nonumber\\
&-A_4^{\rm tree}(p_1,p_2,p_3,-P)\frac{1}{s_{123}} {\cal I}_4^{\rm 1-loop}(\ell,P,p_4,p_5,p_6),
\label{I6fact123}
\end{align}
where $P=p_1{+}p_2{+}p_3$ is on-shell. Tadpole integrands have again 5 degrees of freedom in the numerator (\ref{ansatz}) while there is only one numerical constant $\beta_{n,0}$. However, this time there are no terms that would vanish in all soft limits $p_k\rightarrow0$ -- this is only possible for four-point kinematics. Therefore, imposing the vanishing soft limit in external legs already fixes the soft integrand ${\cal I}_n^{\rm S}$ completely for $n \ge 4$.

%We remark that the integrand obtained from Feynman diagrams does not satisfy (\ref{facttree}) for $n\ge 6$, while the soft integrand proposed here manifestly does.

%%%%%%%%%%%%%%%%%
\section{Single cut}
%%%%%%%%%%%%%%%%%

Let us now look more closely at the single cut of the one-loop integrand. While this is not a forward limit of the tree-level amplitude, it is still a very interesting object which is calculable. At four points (\ref{cut2}) this function $B_6(p_1,p_2,p_3,p_4,-\ell,\ell)$ is equal to
\begin{equation}
    B_6 = 2s_{12} + \ell_1^2 - \ell_{12}^2 + \ell_{123}^2 - \frac{(s_{12} + \ell_1^2)(s_{12} + \ell_{123}^2)}{\ell_{12}^2}. \label{B6fixed}
\end{equation}
This function exhibits a tree-level factorization on the pole $\ell_{12}^2 =0$,
%
%\begin{equation}
%    B_6 \xrightarrow[]{\ell_{12}^2= 0} -\frac{A_4^{\rm tree}(\ell,p_1,p_2,-P)A_4^{\rm tree}(P,p_3,p_4,-\ell)}{\ell_{12}^2},
%\end{equation}
%
\begin{equation}
    B_6 \overset{\ell_{12}^2=0}{\longrightarrow} -\frac{A_4^{\rm tree}(\ell,p_1,p_2,-P)A_4^{\rm tree}(P,p_3,p_4,-\ell)}{\ell_{12}^2},
\end{equation}
with $P=\ell_{12}$, while it lacks all other poles that would be present in the (divergent forward limit of the) tree-level amplitude $A_6(p_1,p_2,p_3,p_4,-\ell,\ell)$. 
%The on-shell function $B_6$ has also a vanishing soft limit for $\ell\rightarrow0$, $p_2\rightarrow0$ and $p_3\rightarrow0$ while it does not vanish for $p_1\rightarrow0$ or $p_4\rightarrow0$. 

This generalizes for higher $n$. We can define a single cut, $\ell^2=0$, of the loop integrand ${\cal I}_n^{\rm 1-loop}$ to be the function $B_{n{+}2}(p_1,{\dots},p_n,-\ell,\ell)$. This function factorizes on the pole $\ell_{12{\dots}m}^2=0$ as
\begin{equation}
   -\frac{A^{\rm tree}_{n_L}(\ell,p_1,{\Ddots},p_m,-P)A^{\rm tree}_{n_R}(P,p_{m{+}1},{\Ddots},p_n,-\ell)}{\ell_{12{\dots}m}^2},\!\! \label{B1}
\end{equation}
where $n_L = m+2$, $n_R = n-2+m$ for $m=2,4,\dots,n{-}2$ and $P=\ell_{12\dots m}$. At the same time, $B_{n{+}2}$ vanishes in the soft limits
\vspace{-0.2cm}
\begin{equation}
 \lim_{\ell\rightarrow0} B_{n{+}2} = 0,\quad \lim_{p_k\rightarrow0} B_{n{+}2} = 0, \label{B2}
 \vspace{-0.2cm}
\end{equation}
for $k=2,{\dots},n{-}1$. The on-shell function $B_{n{+}2}$ only fails to vanish in the soft limit for $p_1\rightarrow0$ and $p_n\rightarrow0$. The conditions (\ref{B1}) and (\ref{B2}) fix $B_{n{+}2}$ completely, and we can reconstruct it using tree-level recursion relations using the shift (\ref{shift}) together with shifting (on-shell) $\widehat{\ell} = \ell(1-az)$.
%
%\begin{equation}
%    \widehat{\ell} = \ell(1-az),
%\end{equation}
%
Note that $a$ is unconstrained as it does not affect momentum conservation. 
%(cancels in $\ell+(-\ell)$ combination). 

The Cauchy formula (\ref{GRT}) then takes the form
\begin{equation}
    \oint \frac{dz}{z}\frac{B_{n{+}2}(z)}{(1-za)F(z)} = 0,\label{recur2}
\end{equation}
and we can reconstruct $B_{n{+}2}\equiv B_{n{+}2}(z=0)$ from factorizations on $\ell_{12{\dots}m}^2=0$. 

Note that here we provide a construction for the single cut $\ell^2=0$ but a completely analogous procedure works for all other $n{-}1$ single cuts. The corresponding on-shell functions are related to $B_{n+2}(p_1,\Ddots,p_n,-\ell,\ell)$ by cyclic shifts. For example, at six points, a single cut $\ell_{12}^2=0$ gives $B_6\equiv B_6(p_3,p_4,p_1,p_2,-\ell_{12},\ell_{12})$,
\begin{equation}
    B_6 = 2s_{12} + \ell_{123}^2 - \ell^2 + \ell_1^2 - \frac{(s_{12}+\ell_{123}^2)(s_{12}+\ell_1^2)}{\ell^2},
\end{equation}
which factorizes into tree amplitudes on the pole $\ell^2=0$ and vanishes for soft limits $\ell_{12}\rightarrow0$, $p_1\rightarrow0$ and $p_4\rightarrow0$.

%%%%%%%%%%%%%%%%%
\section{Loop recursion relations}
%%%%%%%%%%%%%%%%%

Now we are ready to formulate recursion relations for one-loop soft integrands in the planar (large $N$) limit of $SU(N)$ NLSM. The $n$-point one-loop integrand ${\cal I}_n^{\rm 1-loop}$ has two types of poles:

\begin{itemize}
    \item Tree-level poles $s_{i{\dots}j}=0$: the integrand factorizes into a product of a lower point integrand and a tree-level amplitude (\ref{facttree}). 
    \item Single cuts $\ell_{1{\dots}j}^2=0$: evaluate to $-B_{n{+}2}$ functions. 
\end{itemize}

We proceed to shift the external momenta using a tree-level shift (\ref{shift}). The integrand ${\cal I}_n^{\rm 1-loop}$ now depends on $z$ and the shifted integrand ${\cal I}_n^{\rm S}(z)$ vanishes for $z_k=1/a_k$ and evaluates to known functions on both tree-level and loop poles. Then we use the same Cauchy formula (\ref{GRT}), now for ${\cal I}_n^{\rm 1-loop}$, and evaluate the pole at $z=0$ as 
\begin{align}
\begin{split}
    {\cal I}_n^{\rm 1-loop} &= \sum_{I,\pm} \mathop{\mathrm{Res}}_{z=z_I^{\pm}} \frac{{\cal I}_L(z)A_R(z) + A_L(z) {\cal{I}}_R(z)}{zP^2_I(z)F(z)}  \\
    &+ \sum_{m,\pm} \mathop{\mathrm{Res}}_{z=z_m^{\pm}} \frac{B_{n{+}2}(z)}{z\, l_{1{\dots}m}^2(z)F(z)}. \label{recur1}
\end{split}
\end{align}
In the first term we sum over tree-level poles $P_I^2(z_I^{\pm})=0$ and in the second term over single cuts $\ell_{1{\dots}m}^2(z_m^{\pm})=0$. 

To give a simple example, we reconstruct the soft integrand (\ref{I4s}) using (\ref{recur1}). At four points there are no tree-level poles and the integrand is determined by the on-shell function (\ref{B6fixed}) alone. We deform the external momenta as in (\ref{shift}) and shift $\hat{\ell} = \ell + zq$, with some auxiliary vector $q$, $q^2 \neq 0$. The latter shift is needed only for $n=4$ to avoid poles at $z\!=\!\infty$.  Evaluating (\ref{recur1}) we obtain
\begin{align}
\begin{split}
    {\cal I}_4^{\rm S} &= \!\!\!\sum_{m=1,\pm}^{4} \!\!\frac{B_6(\hat{p}_{1+m},\Ddots,\hat{p}_{4+m},-\hat{\ell}_{1\dots m},\hat{\ell}_{1\dots m};z_m^\pm)}{(z_m^\pm/z_m^\mp -1)\,\ell_{1\dots m}^2F(z_m^\pm)},
    \label{I4sRec}
\end{split}
\end{align}
where the two solutions to $\ell_{1{\dots}m}^2(z_m^{\pm})=0$ are $z_m^\pm=$
\begin{align}
     \frac{-\ell_{1\dots m}\!\cdot\! Q_{1\dots m} \pm\! \sqrt{(\ell_{1\dots m}\!\cdot\! Q_{1\dots m})^2 - \ell_{1\dots m}^2Q_{1\dots m}^2}}{Q_{1\dots m}^2}.
\end{align}
For a given shift the vectors $Q_{1\dots m}$ are defined by $\hat{\ell}_{1\dots m}(z) = \ell_{1\dots m} + zQ_{1\dots m}$. Concretely, for the shift used to compute (\ref{I4sRec}) and e.g. $m\!=\!2$ we get $\hat{\ell}_{12}(z)=\ell_{12}+zQ_{12}$ with $Q_{12}=q+q_1-a_2p_2$, and similar for all other $m$. Note that the result (\ref{I4sRec}) does not depend on the shift parameters and agrees with (\ref{I4s}). 

%%%%%%%%%%%%%%%%%
\section{Double soft limit}
%%%%%%%%%%%%%%%%%

Flavor-ordered amplitudes in the NLSM are known to satisfy a recursion relation when the soft limit of two external momenta is taken simultaneously \cite{Kampf:2013vha},
\begin{equation}
    \lim_{t\to 0} A_n(tp_i,tp_j,\lbrace p_k\rbrace) = \Pi_{i,j} A_{n-2}(\lbrace p_k\rbrace),
    \label{DSamp}
\end{equation}
where $1\le i < j \le n$, and $\lbrace p_k\rbrace$, $k\neq i$, $k\neq j$, denotes the dependence on the remaining momenta. The double soft factor $\Pi_{i,j}$ equals
\begin{equation}
    \frac{\delta_{j,i+1}}{2} \left( \frac{p_{i+2}\cdot (p_i-p_{i+1})}{p_{i+2}\cdot (p_i+p_{i+1})} - \frac{p_{i-1}\cdot (p_i-p_{i+1})}{p_{i-1}\cdot (p_i+p_{i+1})} \right).
    \label{DSfact}
\end{equation}
We saw that a soft integrand is uniquely determined by a choice of coefficients $\alpha_1$,$\alpha_2$ in (\ref{cut2}). One way to fix these constants was the prescription (\ref{b6Sft}). Alternatively we can ask if there exists a choice that manifests the double soft condition at the integrand level,
\begin{equation}
 \lim_{t\to 0} \mathcal{I}_n(tp_i,,tp_j,\lbrace\ell, p_k\rbrace) \overset{!}{=} \Pi_{i,j} \mathcal{I}_{n-2}(\lbrace \ell, p_k \rbrace).
 \label{DSint}
\end{equation}
Remarkably, the answer turns out affirmative. At four points the soft integrand ${\cal I}_4^{\rm S}(\alpha_1,\alpha_2)$ vanishes identically in all double soft limits and for all $\alpha_1$,$\alpha_2$. However, at $n=6$ points and up, (\ref{DSint}) represents a non-trivial constraint forcing $\alpha_1=-2$ and $\alpha_2=1$. The corresponding four-point {\it double soft integrand} thus takes the form
\begin{equation}
    {\cal I}_4^{\rm DS} = \frac{1}{2} + \frac{n_4^{\rm DS}}{\ell^2} +\frac{n_{42}}{\ell^2\ell_{12}^2} + \text{cyc},
    \label{I4ds}
\end{equation}
now with $n_4^{\rm DS}= -2s_{12} + s_{23} - \ell_1^2 + \ell_{12}^2 - \ell_{123}^2$.

Equivalently we can reformulate the condition (\ref{DSint}) as a statement about the soft limit $\ell \to 0$ of the six-point on-shell function analogous to (\ref{b6Sft}),
\begin{align}
 \lim_{\ell\rightarrow0} B_6 = -(\alpha_1+2)s_{12} - \alpha_2 s_{23} \overset{!}{=} -s_{23}\,, \label{B6DSCond}
\end{align}
which does not require us to go beyond $n=4$ to unambiguously fix the double soft integrand. 

%As noted before, for higher points we only need to impose the cuts and Adler zero on external legs to fix the integrand uniquely. 
The same recursion relations for $B_{n{+}2}$ (\ref{recur2}) and ${\cal I}_n^{\rm 1-loop}$ (\ref{recur1}) can then be used to construct the double soft integrand ${\cal I}^{DS}_n$. The only difference is the seed: the single cut function $B_6$.  
We can see that $\alpha_1=-2$ in both (\ref{b6Sft}) and (\ref{B6DSCond}), while the coefficient $\alpha_2$ is different. In fact, $\alpha_1=-2$ is required by on-shell constructability of $B_6$ (see \cite{progress} for more details), but $\alpha_2$ is unfixed and can be set to any value. Thus, the one-parametric family of integrands ${\cal I}^{\rm 1-loop}_n(\alpha_2)$ satisfies all cuts and the Adler zero in external legs and can be reconstructed using the recursion (\ref{recur1}). The integrands depend on $\alpha_2$ because the seed functions in the recursion relations do. Here we point out two special values $\alpha_2=0,1$ for which the single cut and integrand exhibit additional soft behavior respectively. 
%Which of them is more fundamental and does survive to higher loops is left for future work.

%%%%%%%%%%%%%%%%%
\section{Conclusion and Outlook}
%%%%%%%%%%%%%%%%%

In this letter we constructed the one-loop planar integrand in the $SU(N)$ non-linear sigma model for all multiplicities. Apart from satisfying the standard cuts it satisfies an Adler zero in all external momenta. These constraints lead to a one-parametric family of {\it soft integrands} with two special cases which have either the vanishing soft limit on single cuts (\ref{b6Sft}) or directly exhibit the double soft limit (\ref{DSint}). We further formulated loop-level soft recursion relations to reconstruct the soft integrands from their cuts. 
In the upcoming work \cite{progress} we will show how our construction extends to two-loops and discuss the generalization to other large $N$ theories of Goldstone bosons. Our work is a first step to systematically study the loop integrands of theories with special soft limits. In the case of ${\cal N}=4$ SYM similar considerations lead to the geometric construction of the perturbative S-matrix in terms of the positive Grassmannian \cite{Arkani-Hamed:2012zlh} and the Amplituhedron \cite{Arkani-Hamed:2013jha, Arkani-Hamed:2017vfh}. The goal of our program is to find analogous geometric structures for theories with special soft limits.
%

%\vspace{-0.5cm}
{\it Acknowledgements}: This work was supported by the projects GA\v{C}R 21-26574S and GAUK-327422. J.T. is supported by the DOE grant No. DE-SC0009999 and by the funds of University of California.

\end{document}